\documentstyle[12pt,bezier]{article}
\newcommand{\rf}[1]{(\ref{#1})}
\newcommand{\beq}{\begin{equation}}
\newcommand{\eeq}{\end{equation}}
\newcommand{\bea}{\begin{eqnarray}}
\newcommand{\eea}{\end{eqnarray}}
\newcommand{\e}{{\rm e}}
\newcommand{\g}{\gamma}

\renewcommand{\b}{\beta}
\renewcommand{\a}{\alpha}
\newcommand{\n}{\nu}
\newcommand{\m}{\mu}
\newcommand{\del}{\delta}

\newcommand{\oh}{\frac{1}{2}}
\newcommand{\oq}{\frac{1}{4}}
\newcommand{\tq}{\frac{3}{4}}

\newcommand{\ra}{\right\rangle}
\newcommand{\la}{\left\langle}
\newcommand{\dm}{\triangle \m}

\newcommand{\smr}{\sqrt{\m_r}}

\newcommand{\hx}{\hat{x}}

\newcommand{\prt}{\partial}
\newcommand{\intx}{\oint_{C_x} \frac{dx}{2\pi {\rm i}x}}
\newcommand{\inty}{\oint_{C_y} \frac{dy}{2\pi {\rm i}y}}

\newcommand{\cT}{{\cal T}}

\newcommand{\cO}{{\cal O}}

\newcommand{\cF}{{\cal F}}

\newcommand{\nn}{\nonumber \\}

\begin{document}
\topmargin 0pt
\oddsidemargin 5mm
\headheight 0pt
\headsep 0pt
\topskip 9mm

\hfill    NBI-HE-95-01

\hfill January 1995

\begin{center}
\vspace{24pt}
{\large \bf Scaling in quantum gravity}

\vspace{24pt}

{\sl J. Ambj\o rn } and {\sl Y. Watabiki}

\vspace{6pt}

 The Niels Bohr Institute\\
Blegdamsvej 17, DK-2100 Copenhagen \O , Denmark\\

\end{center}
\vspace{24pt}

\vfill

\begin{center}
{\bf Abstract}
\end{center}

\vspace{12pt}

\noindent
The 2-point function is the natural object
in quantum gravity for extracting critical behavior:
The exponential fall off of the 2-point function with geodesic
distance determines the fractal dimension $d_H$ of space-time.
The integral of the 2-point function determines the
entropy exponent $\gamma$, i.e. the fractal structure
related to baby universes, while the short distance
behavior of the 2-point function connects $\gamma$ and
$d_H$ by a quantum gravity version of Fisher's scaling
relation. We verify this behavior in the case of
2d gravity by explicit calculation.

\vfill

\newpage

\section{Introduction}
Much has been achieved in our understanding of 2d quantum
gravity, both from the point of view of Liouville theory \cite{ddk}
and from a discretized point of view \cite{david,adf,adfo,kkm}.
However, the simplest
and most fundamental concept in gravity, the concept of
{\it distance}, has only recently been analyzed.
In Liouville theory the tool has been the diffusion
equation for a random walk on the ensemble of 2d manifolds
weighted by the Liouville action \cite{watabiki1}.
In the framework of dynamical
triangulations the tool has been the transfer matrix formalism
developed in \cite{transfer}. In this article we show
that standard scaling relations known from
statistical mechanics follow unambiguously even in
quantum gravity once the geodesic distance is used to
set the length scale in the problem.

\section{Scaling relations}

Let us define 2d quantum gravity as the scaling limit of the
so-called simplicial quantum gravity theory. Simplicial quantum
gravity can be defined in any dimensions, but we will here restrict
ourselves to 2d. The partition function will be given by:
\beq\label{1}
Z(\m,G_E)= \sum_{T \in \cT} \frac{1}{C_T} \e^{-S[T]},
\eeq
where $S[T]$ is the Einstein-Hilbert action:
\beq\label{2}
S[T] = \m N_T -\frac{1}{4\pi G_E} (2 - 2g_T).
\eeq
In eqs. \rf{1} and \rf{2} $\cT$ denotes a suitable class of
triangulations of closed 2-manifolds, $T$ a triangulation in $\cT$,
$N_T$ the number of triangles in $T$, $C_T$ a symmetry factor
and $g_T$ the genus of the manifold.
If we fix the  topology, as we will always do in the following,
we can drop the last term since it is a topological invariant.

It is known that the partition function $Z(\m)$
(for a fixed topology)
has a  critical bare cosmological constant $\m_c$ such that
the continuum limit of \rf{1} should be taken  for $\m \to \m_c$ from above.
Define $\dm \equiv \m-\m_c$, then
\beq\label{2a}
Z (\m) \sim {\rm const.} (\dm)^{2-\g} + {\rm less~singular~terms}
\eeq
For 2d gravity $\g=-1/2$ for spherical topology.
In the following we will consider only triangulations with
spherical topology.

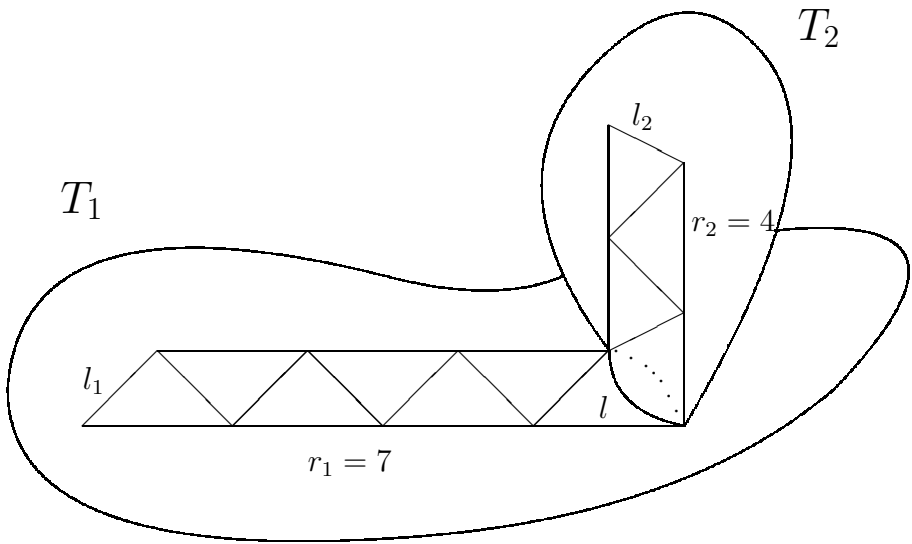
\begin{figure}
\unitlength=1.00mm
\linethickness{0.6pt}
\begin{picture}(143.00,80.00)
\put(20.00,20.00){\line(1,1){10.00}}
\put(30.00,30.00){\line(1,-1){10.00}}
\put(40.00,20.00){\line(-1,0){20.00}}
\put(40.00,20.00){\line(1,1){10.00}}
\put(50.00,30.00){\line(1,-1){10.00}}
\put(60.00,20.00){\line(-1,0){20.00}}
\put(40.00,20.00){\line(1,1){10.00}}
\put(50.00,30.00){\line(1,-1){10.00}}
\put(60.00,20.00){\line(-1,0){20.00}}
\put(60.00,20.00){\line(1,1){10.00}}
\put(70.00,30.00){\line(1,-1){10.00}}
\put(80.00,20.00){\line(-1,0){20.00}}
\put(80.00,20.00){\line(1,1){10.00}}
\put(100.00,20.00){\line(-1,0){20.00}}
\put(80.00,20.00){\line(1,1){10.00}}
\put(100.00,20.00){\line(-1,0){20.00}}
\put(30.00,30.00){\line(1,0){60.00}}
\bezier{72}(90.00,30.00)(90.00,22.00)(100.00,20.00)
\put(100.00,20.00){\line(0,1){15.00}}
\put(100.00,35.00){\line(-2,-1){10.00}}
\put(90.00,30.00){\line(0,1){15.00}}
\put(90.00,45.00){\line(1,-1){10.00}}
\put(100.00,35.00){\line(0,1){20.00}}
\put(100.00,55.00){\line(-1,-1){10.00}}
\put(90.00,45.00){\line(0,1){15.00}}
\put(90.00,60.00){\line(2,-1){10.00}}
\bezier{292}(60.00,5.00)(10.00,2.00)(10.00,25.00)
\bezier{268}(60.00,5.00)(99.00,7.00)(120.00,25.00)
\bezier{216}(90.00,30.00)(72.00,54.00)(90.00,70.00)
\bezier{308}(10.00,25.00)(12.00,52.00)(60.00,40.00)
\bezier{100}(60.00,40.00)(75.00,36.00)(84.00,40.00)
\bezier{256}(112.00,46.00)(143.00,49.00)(120.00,25.00)
\put(91.00,30.00){\circle*{0.00}}
\put(93.00,29.00){\circle*{0.00}}
\put(95.00,28.00){\circle*{0.00}}
\put(96.00,27.00){\circle*{0.00}}
\put(99.00,22.00){\circle*{0.00}}
\put(98.00,24.00){\circle*{0.00}}
\put(97.00,26.00){\circle*{0.00}}
\put(20.00,50.00){\makebox(0,0)[cc]{{\Large $T_1$}}}
\put(118.00,73.00){\makebox(0,0)[cc]{{\Large $T_2$}}}
\put(50.00,15.00){\makebox(0,0)[lc]{$r_1=7$}}
\put(101.00,47.00){\makebox(0,0)[lc]{$r_2=4$}}
\put(23.00,26.00){\makebox(0,0)[rc]{$l_1$}}
\put(96.00,60.00){\makebox(0,0)[rb]{$l_2$}}
\put(90.00,24.00){\makebox(0,0)[rt]{$l$}}
\bezier{244}(110.00,70.00)(122.00,57.00)(100.00,20.00)
\bezier{112}(90.00,70.00)(101.00,80.00)(110.00,70.00)
\end{picture}
\caption[figsc1]{The inequality \rf{6}. Two triangulations with marked
links separated by distances $r_1$ and $r_2$ can be glued together
to a triangulation  where the marked links has a distance $r_1+r_2$
but the same number of triangles by cutting open a marked link
in each of the triangulations to a 2-loop boundary and glue together the
two boundaries.}
\label{figsc1}
\end{figure}

We define the geodesic distance {\it between
two links}\footnote{It is convenient here to use the geodesic distance
between links rather than between vertices. As will be clear later the
results are independent of this definition. We choose it because it is
technically convenient in the analytic calculations.} as
the shortest path of links connecting the two links in the dual
lattice, i.e. the shortest ``triangle-path'' between the two
links on the original triangulation.
Let us by $\cT_2(r)$ denote the ensemble of (spherical)
triangulations with two marked links separated a geodesic
distance $r$ (we assume for the moment that the link length is $a=1$).
We can now define the 2-point function of quantum gravity by
\beq\label{3}
G_\m(r) = \sum_{T \in \cT_2(r)} \e^{-\m N_T}.
\eeq
The 2-point function falls off exponentially for $r \to \infty$.
\beq\label{5}
\lim_{r \to \infty} \frac{-\log G_\m (r)}{r} = m(\dm) \geq 0.
\eeq
This trivial but important relation follows from the
fact that
\beq\label{6}
G_\m(r_1+r_2) \geq  G_\m(r_1)\, G_\m(r_2),
\eeq
simply because each term on the rhs of eq. \rf{6} can
be given an interpretation as a term belonging to the lhs of
eq. \rf{6}. This is illustrated in fig.\,\ref{figsc1}.
Eq. \rf{6} shows that $-\log G_\m (r)$ is sub-additive and this
ensures the existence of the limit \rf{5}. In addition
$m(\dm) \geq 0$ because $G_\m (r)$ is a decreasing function
of $r$. Again this follows from general arguments which allow
us to bound the number of triangulations with $N_T$ triangles and two
marked links separated a distance $r$ in terms of the number
of triangulations with $N_T$ triangles and two marked links
separated a distance $r' < r$. The same kind of arguments lead
to the conclusion that $m'(\dm) >0$ for $\dm > 0$, i.e. $m(\dm)$
is a decreasing function as $\m \to \m_c$.

Similarly we can define $\cT(l_1,l_2;r)$ as the class of triangulations with
an entrance boundary loop $l_1$ (with one marked link)
and an exit boundary loop $l_2$, separated a geodesic distance $r$.
$l_1$ and $l_2$ are the number of links at the entrance boundary loop
and that at the exit boundary loop, respectively.
We say that $l_1$ and $l_2$ are separated by a geodesic distance $r$
if {\it all} links $l \in l_2$ has the geodesic distance $r$ to $l_1$.
Finally  the geodesic distance between a link $l$ and a set of links
(like the loop $l_1$) is the minimum of the  geodesic distances between the
link $l$ and the links in the set.
We define\footnote{Sometimes a different notation is used and
the entrance loop is unmarked while the exit loop is marked. The
difference in these functions will be some trivial factors of
$l$: $l_2G_\m (l_1,l_2;r) = l_1 G'_\m(l_1,l_2;r)$ where $G'$ denotes
the 2-loop function where the exit loop is marked.}
\beq\label{4}
G_\m(l_1,l_2;r) = \sum_{T \in \cT(l_1,l_2;r)} \e^{-\m N_T}.
\eeq
The 2-loop functions fall off exponentially for $r \to \infty$.
Again this follows from the subadditivity argument since one has
\cite{transfer,watabiki}:
\beq\label{4a}
G_\m(l_1,l_2;r_1+r_1) = \sum_{l=1}^{\infty}\; G_\m(l_1,l;r_1)G_\m(l,l_2;r_2).
\eeq
It is easy to show, by arguments identical to the ones used
originally for random surfaces on hypercubic lattices \cite{dfj},
that the mass is defined by the exponential decay
of the two-loop function is independent of the length of the
boundary loops and consequently identical to the mass defined by the
2-point function.

The important point is that the ``mass'' $m(\dm)$ dictates the scaling in
quantum gravity. We can view $G_\m(r)$ as the partition function
for universes of linear extension $r$ and in order that this partition
function survives in the continuum limit it is necessary that we have
\beq\label{7}
m(\dm) r = M\, R,
\eeq
where $M$ and $R$ are kept fixed in the continuum limit where
the number of lattice steps $r$ goes to infinity.  There can only be
a continuum limit if $m(\dm) \to 0$ for $\dm \to 0$. In addition
almost all critical properties of quantum gravity can be read off
directly from this function in the scaling limit. This has
already been emphasized in the more general context of higher dimensional
quantum gravity \cite{adj} and more specifically in
two dimensions \cite{adj1} (where an explicit solution was
given for a toy model of branched polymers),
but it is worth to repeat the arguments.
First one would in general expect the exponential decay of $G_\m(r)$ to
be replaced by a power fall off when $m(\dm) =0$, or more precisely
in the region where $1 \ll r \ll 1/m(\dm)$. The behavior $G_\m (r)$ is
is assumed to be:
\bea
G_\m (r) &\sim & \e^{-m(\dm) r}~~~~~~~~~~~~
{\rm for}~~ m(\dm) r \gg 1\label{8} \\
G_\m (r) &\sim & r^{1-\eta} ~~~~~~~~~~~~~~~~~~{\rm for}~~
1 \ll r \ll \frac{1}{m(\dm)}
\label{9} \\
\chi_\m & \equiv & \int \! dr \;G_\m (r)
\sim \frac{\prt^2 Z(\m)}{\prt \m^2} \sim
{\rm const.} \, (\dm)^{-\g}.\label{10}
\eea
These are standard definitions in statistical mechanics. The exponent $\eta$
is called the anomalous scaling dimension since a free propagator in any
space-time dimensions has $\eta =0$. This follows by integrating the usual
free propagator over the angular variables corresponding to a fixed value of
$r$.

Two scaling relations follow directly from the definitions if we assume
that $m(\dm) \to 0$ for $\dm \to 0$ as:
\beq\label{11}
m(\dm) \sim (\dm)^\n.
\eeq
{}From eq. \rf{10} it follows, after differentiating a sufficient number
of times after $\m$ that
\beq\label{12}
\g = \n (2-\eta),
\eeq
a relation known in statistical mechanics as {\it Fisher's scaling relation}.
In that case it will typically be a relation between the spin susceptibility
exponent $\g$, the critical exponent $\n$ of the spin-spin
correlation length and
the anomalous scaling exponent of the spin-spin correlation function.
It is remarkable that it is still valid in quantum gravity.
The other scaling relation is
\beq\label{13}
\n = \frac{1}{d_H},
\eeq
where $d_H$ denotes the (internal) Hausdorff dimension of
the ensemble of random surfaces given by eq. \rf{3}.
To be more precise we define the (internal) Hausdorff dimension
of this ensemble by
\beq\label{14}
\la N \ra_r \sim r^{d_H},~~~~r \to \infty ,~~~m(\dm)r = {\rm const.}
\eeq
where
\beq\label{15}
\la N \ra_r \equiv
\frac{\sum_{T \in \cT_2(r)} N_T\; \e^{-\m N_T}}{\sum_{T \in \cT_2(r)}
\,\e^{-\m N_T}}.
\eeq
It follows from the definitions that:
\beq\label{16}
\la N\ra_r \sim -\frac{1}{G_\m (r)}\;\frac{\prt G_\m (r)}{\prt \m} \sim
m'(\dm) r \sim r^{1/\n}.
\eeq

It is interesting to give a direct physical interpretation
of the short distance behavior of the $G_\m (r)$ as defined
by \rf{9}. In order to do so let us change from the {\it grand canonical
ensemble} given by \rf{3} to the {\it canonical ensemble} defined by
\beq\label{hx1}
G(r,N) = \sum_{T \in \cT_2(r,N)} 1,
\eeq
where $\cT_2(r,N)$ denotes the triangulations which $N$ triangles and
two marked links separated a distance $r$.
$G(r,N)$ is the 2-point function where the number of triangles is
fixed to $N$. For  $r=0$ we have the following
$N$ dependence (for the $r$ dependence see \rf{hx6} below)
\beq\label{hx2}
G(0,N) \sim N^{\g-2}\, \e^{\m_c N}.
\eeq
The reason is that the partition function for a finite volume $N$
is assumed to behave like
\beq\label{hx2a}
Z(N) \sim N^{\g -3} \, \e^{\m_c N}.
\eeq
The 1-point function is for large $N$ proportional to
$N Z(N)$ since it counts the triangulations with one marked
link or triangle or vertex depending on the precise definition,
and for $r=0$ (or just small) there is essntially no difference
between the 1-point function and $G(0,N)$.

$G(r,N)$ is related to $G_\m(r)$ by a (discrete) Laplace transformation:
\beq\label{hx3}
G_\m(r) = \sum_N G(r,N)\, \e^{-\m N}.
\eeq
The {\it long distance behavior} of $G(r,N)$ is determined by the
long distance behavior of $G_\m(r)$. Close to the scaling
limit it follows by direct calculation (e.g. a saddlepoint calculation) that
\bea
G_\m(r) &\sim &\e^{-r\,(\dm)^{1/d_H} }~~~ \Rightarrow   \nn
G(r,N) &\sim & \e^{-c \left(r^{d_H}/N\right)^{\frac{1}{d_H-1}}}\; \e^{\m_c N}
{}~~~~{\rm for}~~~~ r^{d_H} > N,    \label{hx4}
\eea
where $c=(d_H-1)/d_H^{d_H/(d_H-1)}$.

On the other hand the {\it short distance behavior} of $G_\m (r)$
is determined by the short distance behavior of $G(r,N)$ which is
simple. Eqs. \rf{14} and \rf{15} defined the concept of
Hausdorff dimension in the grand canonical ensemble. A
definition in the canonical ensemble would
be: Take $N^{1/d_H} \gg r$ and simply count the  volume
(here number of triangles) of a ``spherical shell'' of thickness 1 and
radius $r$  from a marked link,
sum over all triangulations with one marked link and $N$ triangles, and
divide by the total number of triangulations with one marked link
and $N$ triangles. Call this number $\la n(r)\ra_N$. The Hausdorff
dimension is then defined by
\beq\label{hx5}
\la n(r) \ra_N \sim r^{d_H-1}~~~{\rm for}~~~1 \ll r \ll N^{1/d_H}.
\eeq
It follows from the definitions that we can write
\bea
\la n(r) \ra_N &\sim & \frac{G(r,N)}{G(0,N)}, ~~~~{\rm i.e}~~~\nn
G(r,N)& \sim& r^{d_H-1} N^{\g-2} \e^{\m_c N}~~~{\rm for}~~~
1 \ll r \ll N^{1/d_H}.                            \label{hx6}
\eea
We can finally calculate the short distance behavior of $G_\m(r)$
from eq. \rf{hx3}. From \rf{hx4} the sum is cut off at $N \sim r^{d_H}$.
For $\m \to \m_c$, i.e. $\dm$ small we get:
\beq\label{hx7}
G_\m(r) \sim r^{d_H-1} \sum_{N=1}^{r^{d_H}} N^{\g-2}  \sim r^{\g d_H -1}.
\eeq
This is actually an independent derivation of Fisher's scaling
relation since it shows directly that $\eta = 2-\g d_H$, and it has
the advantage that it gives a physical interpretation of the
anomalous scaling dimension $\eta$. In addition it proves that
the canonical and grand canonical definition of Hausdorff dimension
in fact agrees.

The model of branched polymers ($BP$)
provides us with a simple, but non-trivial
example of the above scenario \cite{adj1}.
Here we will define branched polymers as the sum over all
tree graphs (no loops in the graphs) with certain weights
given to the graphs according to the following definition of the
partition function:
\beq\label{k2}
Z(\m) = \sum_{BP} \frac{1}{C_{BP}}\rho(BP) \; \e^{-\m |BP|},
\eeq
where $|BP|$ is the number of links in a $BP$ and $\m$ is a chemical
potential for the number of links, while
\beq\label{k3}
\rho(BP)= \prod_{i \in BP} f(n_i),
\eeq
where $i$ denotes a vertex, $n_i$ the number of links joining at
vertex $i$ and $f(n_i)$ is non-negative. $f(n_i)$ can be viewed
as the unnormalized branching weight for one link branching into $n_i-1$
links at vertex $i$. Finally $C_{BP}$ is a symmetry factor such
that rooted branched polymers, i.e. polymers with the first link marked,
is counted only once.

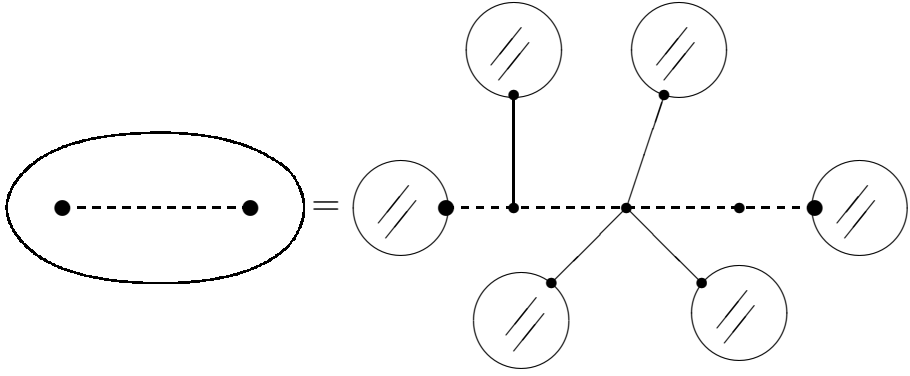
\begin{figure}
\unitlength=1.00mm
\linethickness{0.6pt}
\begin{picture}(127.00,52.00)
\put(15.00,25.00){\circle*{2.00}}
\put(40.00,25.00){\circle*{2.00}}
\put(15.00,25.00){\dashbox{1.00}(25.00,0.00)[cc]{}}
\put(60.00,25.00){\circle{12.00}}
\put(57.00,23.00){\line(4,5){4.00}}
\put(58.00,21.00){\line(4,5){4.00}}
\put(66.00,25.00){\circle*{2.00}}
\put(66.00,25.00){\dashbox{1.00}(49.00,0.00)[cc]{}}
\put(115.00,25.00){\circle*{2.00}}
\put(75.00,25.00){\line(0,1){15.00}}
\put(75.00,40.00){\circle*{1.50}}
\put(75.00,25.00){\circle*{1.50}}
\put(90.00,25.00){\circle*{1.50}}
\put(90.00,25.00){\line(-1,-1){10.00}}
\put(80.00,15.00){\circle*{1.50}}
\put(90.00,25.00){\line(1,-1){10.00}}
\put(100.00,15.00){\circle*{1.50}}
\put(90.00,25.00){\line(1,3){5.00}}
\put(95.00,40.00){\circle*{1.50}}
\put(105.00,25.00){\circle*{1.50}}
\put(75.00,46.00){\circle{12.00}}
\put(72.00,44.00){\line(4,5){4.00}}
\put(73.00,42.00){\line(4,5){4.00}}
\put(97.00,46.00){\circle{12.00}}
\put(94.00,44.00){\line(4,5){4.00}}
\put(95.00,42.00){\line(4,5){4.00}}
\put(76.00,10.00){\circle{12.00}}
\put(74.00,8.00){\line(4,5){4.00}}
\put(75.00,6.00){\line(4,5){4.00}}
\put(105.00,11.00){\circle{12.00}}
\put(102.00,9.00){\line(4,5){4.00}}
\put(103.00,7.00){\line(4,5){4.00}}
\put(121.00,25.00){\circle{12.00}}
\put(118.00,23.00){\line(4,5){4.00}}
\put(119.00,21.00){\line(4,5){4.00}}
\put(50.00,25.00){\makebox(0,0)[cc]{{\large $=$}}}
\bezier{56}(10.00,30.00)(5.00,25.00)(10.00,20.00)
\bezier{52}(45.00,30.00)(49.00,25.00)(45.00,20.00)
\bezier{76}(28.00,35.00)(39.00,35.00)(45.00,30.00)
\bezier{80}(10.00,30.00)(15.00,35.00)(28.00,35.00)
\bezier{80}(10.00,20.00)(15.00,15.00)(28.00,15.00)
\bezier{76}(28.00,15.00)(40.00,15.00)(45.00,20.00)
\end{picture}
\caption[fig3]{The graphical representation of the 2-point
function for branched polymers. The dashed line represents the
unique shortest path between the two marked vertices. The ``blobs''
represent the contribution from all rooted polymers branching
out from a vertex.}
\label{fig3}
\end{figure}

This model can be solve \cite{adfo,adj1}.
It has a critical point $\m_c$ (depending on $f$)
and close to the critical point we have:
\beq\label{k4}
Z''(\m) \sim (\dm)^{-1/2},~~~~\dm \equiv \m -\m_c,
\eeq
i.e. $\g =1/2$ for branched polymers.
On the branched polymers we define the ``geodesic distance'' between
two vertices as the shortest link path, which is unique since we
consider tree-graphs. The graphical representation of the
2-point function is show in fig.\,\ref{fig3}.
Had it not been for the ability to branch, the 2-point function
would simply be
\beq\label{k5}
G_\m(r) = \e^{-\m r}.
\eeq
However, the insertion of 1-point functions at any vertex leads
to a non-analytic coupling constant renormalization and the
result is changed to \cite{adj1}
\beq\label{k6}
G_\m (r) = {\rm const.}\; \e^{-\kappa\, r\sqrt{\dm}}~~~
{\rm for}~~~\dm \to 0,
\eeq
where $\kappa$ is some positive constants depending on $f$.
We can now find $G(r,N)$ by an inverse Laplace transformation:
\beq\label{k7}
G(r,N) = {\rm const.} \, N^{-3/2} r \,\e^{-\kappa^2r^2/4N}.
\eeq
We confirm from this explicit expression
that the (internal) Hausdorff dimension of
branched polymers is 2 (like a smooth surface !)
and that $\g = 1/2$ since the
prefactor of $G(r,N)$ for small $r$ should be $N^{\g-2} r^{d_H-1}$.

It should be emphasized again that these definitions and scaling
relations are  valid for simplicial gravity in three and
four dimensions as defined in \cite{adj,aj,am1}.
In the rest of this paper we will study  how they are realized in
2d simplicial quantum gravity where the exact solution can be found.

\section{The 2-point function}

Let us first define the generating function for 2-loop amplitudes \rf{4} by:
\beq\label{x1}
G_\m(x,y;r) = \sum_{l_1,l_2 =1}^\infty
x^{l_1} y^{l_2} G_\m(l_1,l_2;r) .
\eeq
We can reconstruct $G_\m(l_1,l_2;r)$ by
\beq\label{x2}
G_\m(l_1,l_2;r) = \intx \, x^{-l_1} \inty\, y^{-l_2} \; G_\m(x,y;r),
\eeq
where the contours $C_x$ and $C_y$ surround the origin and avoid
the cuts of $G_\m(x,y;r)$.
The fundamental composition law \rf{4a} reads:
\beq\label{x3}
G_\m(x,y;r_1+r_2) = \oint_C \frac{dz}{2\pi {\rm i} \,z} \;
G_\m(x,\frac{1}{z};r_1)\, G_\m(z,y;r_2).
\eeq
The boundary condition to be imposed is that
\beq\label{x4}
G_\m (l_1,l_2;r=0) = \del_{l_1,l_2}~~~~~{\rm or}~~~~~
G_\m(x,y;r=0)=\frac{xy}{1-xy}.
\eeq

The important insight obtained in \cite{transfer,watabiki}
is that the 2-loop function satisfies a simple  differential equation.
Using the so-called peeling decomposition defined in \cite{watabiki}
the differential equation has the form \cite{watabiki},
\beq\label{x5}
\frac{\prt }{\prt r} G_\m(x,y;r) = x\frac{\prt}{\prt x}
\left(2x^2 f_\m(x) G_\m(x,y;r)\right),
\eeq
which gives the same differential equation in the limit $\dm \rightarrow 0$
as was obtained by combinatorial arguments in \cite{transfer}.

Let $F_\m(x)$ denote
the generating functional for 1-loop functions with one marked link:
\beq\label{x6}
F_\m(x) = \sum_l x^l F_\m(l) .
\eeq
It is well known that
\bea
F_\m(x) &=& \oh \left( \frac{1}{x^2} - \frac{g}{x^3} \right) + f_\m(x),
{}~~~~g\equiv\e^{-\m},\label{x7} \\
f_\m(x) &=&
\frac{g}{2x} (\frac{1}{x}-c_2)
\sqrt{(\frac{1}{x}-c_1)(\frac{1}{x}-c_0)}, \label{x7a}
\eea
where $c_0 < 0 < c_1 < c_2$ as long as $\m > \m_c$. The only thing we need
to know is that {\it at} the critical point $\m_c$ we have $c_2=c_1$,
(which we denote $1/x_c$) and away from the critical point
\bea
c_2(\m) &=& 1/x_c + \frac{\a}{2}\sqrt{\dm} + \cO(\dm),\label{xb7}\\
c_1(\m) &=& 1/x_c - \a \sqrt{\dm} + \cO (\dm), \label{xb7a} \\
c_0(\m) &=& c_0(\m_c) + \cO(\dm).\label{xb7b}
\eea
Here $\a$ is a positive constant\footnote{The values of the constants
which enter are as follows: $\e^{-\m_c} = g_c = 1/(2{\cdot}3^{\tq})$,
$x_c = \oh (3^{\oq}-3^{-\oq})$,
$c_0(\m_c) = 3^{\frac{3}{4}}(1-3^{\oh})$ and $\a = 4 \cdot 3^{-\oq}$.}
of order $\cO(1)$ and the scaling limit is obtained
when $x = x_c - \cO(\sqrt{\dm})$. In this region it is seen that
\beq\label{x8}
f_\m(x) \sim (\dm)^{3/4}
\eeq
and this is the reason the difference equation originating
from \rf{x3} with $r_1=1$ can be replaced with a differential
equation for $\dm \to 0$ even if $r$ is discrete.

The solution to \rf{x4} and \rf{x5} is:
\beq\label{x9}
G_\m(x,y;r) =   \frac{\hx^2 f_\m(\hx)}{x^2 f_\m(x)} \; \frac{\hx y}{1-\hx y}.
\eeq
Here $\hx(x,r)$ is the solution to the characteristic equation
of the partial differential equation \rf{x5}.
The integral of the characteristic equation is
\beq\label{x10}
 r = \int_x^{\hx(x,r)} \frac{dx'}{ 2 x'^3 f_\m(x')} =
\left[ \frac{1}{\del_0}\,
\sinh^{-1}\sqrt{\frac{\del_1}{1-c_2 x'}-\del_2} \;
\right]^{x'=\hx(x,r)}_{x'=x} ,
\eeq
and this expression can be by inverted to give:
\beq\label{y2}
\hx(x,r)= \frac{1}{c_2} \, - \, \frac{\del_1}{c_2}\;
\frac{1}{\sinh^{2}(\del_0 r + \sinh^{-1}\sqrt{\frac{\del_1}{1-c_2 x}-\del_2)}
+ \del_2} ,
\eeq
where $\del_0$, $\del_1$ and $\del_2$ are all positive and defined by
\bea
\del_0 &=& \frac{g}{2} \sqrt{(c_2-c_1)(c_2-c_0)}
        =  \cO((\dm)^{\oq}),\label{x11a} \\
\del_1 &=& \frac{(c_2-c_1)(c_2-c_0)}{c_2(c_1-c_0)}
        =  \cO(\sqrt{\dm}),\label{x11b} \\
\del_2 &=& - \, \frac{c_0(c_2-c_1)}{c_2(c_1-c_0)}
        = \cO(\sqrt{\dm}).\label{x11c}
\eea
It is readily checked that $\hx \to 1/c_2$ for $r \to \infty$ and
$\hx(x,r=0)=x$.
In principle we can calculate $G_\m(l_1,l_2;r)$ from eqs. \rf{x2}, \rf{x9}
and \rf{y2}. Let us only here verify that the
exponential decay of $G_\m(l_1,l_2;r)$ is independent of $l_1$ and $l_2$.
For $r \to \infty$ one gets
\beq\label{y4}
G_\m(l_1,l_2;r)= {\rm const.\,}  \del_0 \del_1 \, \e^{-2\del_0 r}
+ \cO(\e^{-4 \del_0 r}),
\eeq
where {\it const.} is a function of order $\cO(1)$ which depends
on $c_0, c_1, c_2, l_1$ and $l_2$.

\begin{figure}
\unitlength=1.00mm
\linethickness{0.6pt}
\begin{picture}(140.00,70.00)
\put(55.00,30.00){\line(0,1){10.00}}
\put(55.00,40.00){\line(1,-1){10.00}}
\put(65.00,30.00){\line(0,1){10.00}}
\put(65.00,40.00){\line(1,-1){10.00}}
\put(75.00,30.00){\line(0,1){10.00}}
\put(75.00,40.00){\line(1,-1){10.00}}
\put(85.00,30.00){\line(0,1){10.00}}
\put(85.00,40.00){\line(1,-1){10.00}}
\put(95.00,30.00){\line(0,1){10.00}}
\put(95.00,40.00){\line(1,-1){10.00}}
\put(105.00,30.00){\line(0,1){10.00}}
\put(55.00,40.00){\line(1,0){50.00}}
\put(55.00,30.00){\line(1,0){50.00}}
\bezier{56}(105.00,40.00)(105.00,49.00)(100.00,50.00)
\bezier{52}(105.00,30.00)(105.00,22.00)(100.00,20.00)
\bezier{344}(40.00,35.00)(42.00,68.00)(95.00,70.00)
\bezier{368}(40.00,35.00)(40.00,8.00)(105.00,5.00)
\bezier{260}(95.00,70.00)(128.00,70.00)(100.00,55.00)
\bezier{48}(100.00,55.00)(95.00,50.00)(100.00,50.00)
\bezier{80}(115.00,15.00)(107.00,17.00)(95.00,15.00)
\bezier{72}(95.00,15.00)(89.00,16.00)(100.00,20.00)
\put(100.00,50.00){\circle*{0.00}}
\put(99.00,49.00){\circle*{0.00}}
\put(98.00,48.00){\circle*{0.00}}
\put(97.00,46.00){\circle*{0.00}}
\put(96.00,44.00){\circle*{0.00}}
\put(96.00,42.00){\circle*{0.00}}
\put(96.00,28.00){\circle*{0.00}}
\put(96.00,26.00){\circle*{0.00}}
\put(97.00,24.00){\circle*{0.00}}
\put(98.00,22.00){\circle*{0.00}}
\put(99.00,21.00){\circle*{0.00}}
\put(100.00,20.00){\circle*{0.00}}
\bezier{188}(100.00,50.00)(127.00,64.00)(135.00,50.00)
\bezier{88}(135.00,50.00)(140.00,40.00)(135.00,30.00)
\bezier{172}(135.00,30.00)(131.00,19.00)(100.00,20.00)
\put(52.00,34.00){\makebox(0,0)[cc]{$l_1$}}
\put(108.00,35.00){\makebox(0,0)[lc]{$l$}}
\put(70.00,27.00){\makebox(0,0)[lt]{$r=10$}}
\put(82.00,53.00){\makebox(0,0)[rc]{{\large $G_\mu(l_1=1,l_2)$}}}
\put(135.00,60.00){\makebox(0,0)[rc]{{\large $l_2 \;G_\mu (l_2)$}}}
\put(118.00,45.00){\vector(-1,0){13.00}}
\put(121.00,45.00){\makebox(0,0)[rc]{$l_2$}}
\put(32.00,34.00){\makebox(0,0)[rc]{{\Huge $\Sigma$}{\LARGE $_{l_2}$}}}
\bezier{220}(105.00,5.00)(137.00,7.00)(115.00,15.00)
\end{picture}
\caption[fig2]{The 2-point function represented as a summation over 2-loop
functions times 1-loop functions.}
\label{fig2}
\end{figure}
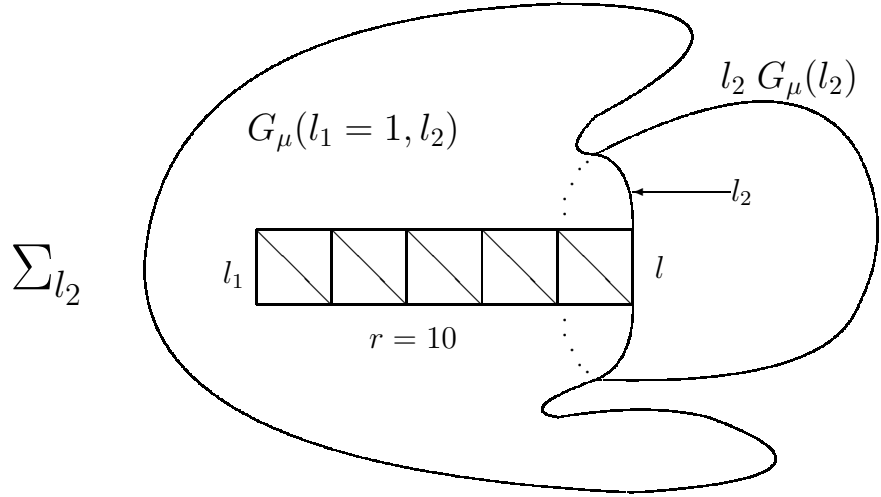

We can express the 2-point function $G_\m(r)$ in terms of the
2-loop function and the 1-loop function. Let us consider a marked
link. For a given triangulation we can systematically work
our way out to the links having a distance $r$ from the marked link
by peeling off layers of triangles having the distances $1,2,\ldots,r$
to the marked link. After $r$ steps we have a boundary consisting
of a number of disconnected boundary loops, all with a distance $r$ to the
marked link. One of these is the exit loop described by the 2-loop function
and we get the 2-point function by closing the exit loop of length $l_2$
by multiplying the 2-loop function $G_\m(l_1=1,l_2;r)$ by the 1-loop function
$F_\m(l_2)$\footnote{To be more precise we have to multiply the 2-loop function
$G_\m(l_1,l_2;r)$ by $l_2$ since the exit loop is unmarked and we
can glue the marked one-loop cap to it in $l_2$ ways.}
 and the perform the sum over $l_2$, i.e., as shown in fig.\,\ref{fig2},
\bea
\lefteqn{G_\m(r) = \sum_{l_2=1}^\infty G_\m(l_1=1,l_2;r)\, l_2 F_\m(l_2)}\nn
 & &= \frac{\prt}{\prt x} \inty \, G_\m(x,\frac{1}{y};r)
      y \frac{\prt}{\prt y} F_\m(y) \Big|_{x=0} \nn
 & &= \frac{\prt}{\prt x} F_\m(\hx) \Big|_{x=0}
    = \frac{1}{g} \frac{\prt}{\prt r} F_\m(\hx) \Big|_{x=0} .
\label{x12}
\eea
As long as $c_2-c_1$ is small
and $r$ is larger than a few lattice spacings, we get:
\beq\label{x14}
G_\m (r) = {\rm const.} \del_0 \del_1
\frac{\cosh( \del_0 r)}{\sinh^3 (\del_0 r)}
\left(1+\cO(\del_0)\right).
\eeq
Formula \rf{x14} shows
how to take the scaling limit: Let us return to the original formulation
and write in the limit $\dm \to 0$:
\beq\label{x15}
G_\m(r) = {\rm const.} \; (\dm)^{3/4}
\frac{\cosh \left[(\dm)^{\oq} \b r\right]}{\sinh^3
\left[(\dm)^{\oq} \b r\right]},
\eeq
where {\it const.} and $\b$ are positive constants
of order $\cO(1)$ ($\b = \sqrt{6} g_c$).

We conclude the following:
\begin{enumerate}
\item $G_\m(r)$  falls of like $\e^{-2(\dm)^{\oq} \b r}$ for $r \to \infty$,
i.e. the critical exponent $\n = \oq$ and the Hausdorff dimension
$d_H =4$.
\item $G_\m(r)$ behaves like $r^{-3}$ for $1 \ll r \ll \dm^{-\oq}$, i.e.
the scaling exponent $\eta = 4$.
\item From Fisher's scaling relation we get $\g = \n (2-\eta) = -1/2$.
This well known result can of course also be derived directly from
\beq\label{yy1}
\chi_\m =  \sum_{r=1}^\infty G_\m(r) = {\rm const.} -c^2 (\dm)^{\oh} + \cdots
\eeq
by use of \rf{x15}, but it should be clear that the explicit calculation
in \rf{yy1} is nothing but a specific example of the general calculation
used in proving Fisher's scaling relation. What is somewhat
unusual compared to ordinary statistical systems is that the
anomalous scaling dimension $\eta >2$. $\eta =0$ is the ordinary
free field result, while $\eta =2$ is the infinite temperature limit,
and for statistical systems we expect $\eta < 2$. A final comment
to \rf{yy1} is that the contant in front of $(\dm)^{\oh}$ is negative,
as indicated by the notation. This has a direct physical interpretation:
$G_\m(r)$ is by definition positive and the same is the case for $\chi_\m$.
However, since $\g = -1/2$ it follows that
$\chi$ will not be divergent at the critical point $\m_c$. But
\beq\label{yy2}
\tilde{\chi}_\m \equiv -\frac{d \chi_\m}{d\m} \sim \frac{c^2}{(\dm)^{\oh}}
+ \cdots
\eeq
{\it is} divergent for $\m \to \m_c$ and {\it has} to be positive
since  it, close to the scaling limit,
has the interpretation as the sum over all triangulations
with three marked links.
\item Any 2-loop function
$G_\m(l_1,l_2;r)$ has the same behavior as $G_\m(r)$ as long
as $l_1,l_2$ stay finite as $\dm \to 0$.
\end{enumerate}

\noindent
It is clear that we could have taken the continuum limit almost at
any point in the above calculations, and in fact already in the
basic equation \rf{x5} by the substitution:
\beq\label{x17}
\frac{1}{x} = \frac{1}{x_c} + \a \xi a,~~~R = \b r \sqrt{a} ,~~~~\dm = \m_r
a^2,
\eeq
\beq\label{x18}
f_\m(x) \sim a^{3/2}\cF_{\m_r}(\xi)+ \cO(a^{5/2}),~~~~
\cF_{\m_r}(\xi) = (\xi - \oh \smr)\sqrt{\xi+\smr},
\eeq
where $\cF_{\m_r}(\xi)$ is the universal disk-amplitude \cite{david3,am}.
The reason we kept the discretized version thoughout the calculation
was to avoid any ambiguity in going from the 2-loop function to
the 2-point function. Properties of the continuum 2-loop function
have already been studied \cite{transfer,watabiki,gk}, but in the
continuum version the length of the bounday loops $l_i$ is
already taken to infinity by $L = l a$, $a \to 0$, $l\to  \infty$, $L$ fixed.
To get the 2-point function we would have to take the limit $L_1 \to 0$
in the continuum version $G_{\m_r}(L_1,L_2;R)$ of $G_\m(l_1,l_2;r)$.
We avoid this ambiguity and can write directly for the continuum
2-point function:
\beq\label{x19}
G_{\m_r}(R) = \lim_{a\to 0} (\sqrt{a})^{\eta-1} G_\m (r) \sim
(\m_r)^{3/4}\frac{\cosh [(\m_r)^{\oq} R]}{\sinh^3 [(\m_r)^{\oq} R]}.
\eeq
The factor in front of $G_\m(r)$ is the usual ``wave function renormalization''
present in the path integral representation of the propagator.
In this way the ``mass'' \rf{7} $M=2 \m_r^{1/4}$, the unusual
power due to $d_H=4$.  Again we find by explicit calculation the
continuum version of \rf{yy1}
\beq\label{yy4}
\chi_{\m_r} = \int_0^\infty  dR\; G_{\m_r} (R) \sim
 \frac{{\rm const.}}{a} - \frac{1}{6} \m_r^{1/2},
\eeq
where the constant in front of $\m_r^{\oh}$ has to be negative for the
reasons mentioned above.

It is interesting to note that the zero of $f_\m(x)$ (or $\cF_{\m_r}(\xi)$)
determines the infinite $r$ limit of the 2-loop (or 2-point) function.
This follows directly from the solution \rf{x10} to
the characteristic equation for \rf{x5}. The distance $r$ can only
diverge if $\hx \to 1/c_2$, the zero of $f_\m(x)$. This zero is usually
uniquely determined by the requirement that the generating functional
$F_\m (x)$ is analytical away from a cut on the real axis and
goes to a constant for $|x| \to 0$. We now see  a direct
physical interpretation: The zero of $f_\m(x)$, or more suggestive:
the pole of $1/f_\m(x)$, determines the mass of the 2-point function.

Recall that for the branched polymer model we found that the
2-point function is $G^{(BP)}_\m (r) \sim \e^{-\kappa \sqrt{\dm}\, r}$,
or introducing continuous variables $\sqrt{\dm} = M a$ and
$R = \kappa r a$:
\beq\label{yy5}
G_M^{(BP)} (R) = \e^{-M R },
\eeq
and compare this to the 2d quantum gravity 2-point function \rf{x19}:
\beq\label{yy6}
G_M (R) = \frac{1}{8} M^3 \frac{\cosh M R/2}{\sinh^3 M R/2}=
 \oh M^3 \sum_{n=1}^\infty n^2\; \e^{-n MR}.
\eeq
While there is only a single mass excitation for the branched polymer
model and it from this point of view seems rather trivial, the
2d gravity model seems to contain an infinite tower of equidistant
mass excitations. Since we get the susceptibility $\chi(\m_r)$
by integrating $G_M(R)$ with respect to $R$ we get the following
formal expression\footnote{The divergence of the sum comes from
the short distance behavior of $G_M(R)$. In the discretized version
of the theory this singular behavior is modified for
$R \sim 1/\sqrt{a}$ and a ``physical cut off'' would be present
such that $$\chi(\m_r) = \oh M^2 \sum_n n \,\e^{-nM\sqrt{a}} =
\frac{1}{2a} - \frac{1}{6} \sqrt{\m_r}$$ in agreement with \rf{yy4}.}
for $\chi(\m_r)$
\beq\label{yy7}
Z''(\m_r) = \chi(\m_r) = \oh M^2 \sum_{n=1}^\infty n ~~
(= -\frac{1}{6} \m_r^{1/2}),
\eeq
where the last equality sign uses $\sum_{n=1}^\infty n = -1/12$.
It agrees of course with the universal part of \rf{yy4} but is interesting
since integration (ignoring non-universal parts of $Z(\m_r)$) leads
to the following representation of $Z(\m_r)$ as a sum over
``mass excitations'':
\beq\label{yy8}
Z(\m_r) = \frac{4}{3} \m_r^{9/4} \sum_{n=1}^\infty nM.
\eeq

\section{discussion}
We have shown that the 2-point function $G_\m(r)$ is a natural
variable which allows us to extract scaling relations in a simple
way. We tested the procedure in 2d quantum gravity which can be solved
analytically and found $\n=1/4$ and $\eta=4$. From the scaling relations
\rf{11} and \rf{12} we conclude that $\g=-1/2$ (which is of course well known)
and $d_H=4$. At first sight it might be surprising that the Hausdorff
dimension is 4, which implies that the geodesic distances scale like
$r\sqrt{a} $ rather than like $a\,r $, where $a$ is the lattice spacing
in the triangulations. However, such non-trivial scaling is unavoidable
if $d_H \neq 2$ and here it has the following simple interpretation:
A boundary of $l$ links will have the discrete length $l$ in lattice
units, but if we view the boundary from the interior of the surface
its true linear extension $r$ will only be $\sqrt{l}$ since the
boundary can be viewed as a random walk from the interior. If we
insist on a continuum limit where we have surfaces with macroscopic boundaries
of length $L = a\,l$ and  ``physical'' area $A=Na^2$, $a$ being the length
unit of the links, such that $L^2 \sim A$,  we are led to
\beq\label{35}
A \sim L^2 \sim a^2 l^2 \sim a^2 r^4 \sim R^4.
\eeq

The above calculation is independent of the fact that we used triangulations
as the underlying discretization. Any resonable distribution of
polygons will give the same results. Once the distribution of polygons
is fixed there is a unique critical point $\m_c$ of the chemical
potential for polygons. The discretized equation will still be given
by \rf{x5}, only will $f(x)$ be a higher order (even infinite order)
polynomium times a square root cut. Close to the critical point
it will still maintain the structure \rf{x7a} as follows from
the general analysis of matrix models \cite{ackm} and the results will
be unchanged for $\dm \to 0$. In particular this shows that we would
have obtained the same results if we used the shortest link length
as geodesic distance, rather than the shortest link length on the dual lattice,
since we could instead perform the summation over $\phi^3$ graphs. These
would be in one-to-one correspondence with the triangulations and our
definition
of geodesic distance on this class of graphs would correspond to the
link-length
definition on the triangulations.

We find formula \rf{yy8} quite interesting. It should be possible to
understand the mass excitations $nM$ in terms of Liouville theory.

It is not clear that the program will work well in the case of
the so-called multicritical matrix models. In these models
the different polygons used in the discretization are not
assigned a positive weight in the summation and the basic inequalities
like \rf{6} are not valid.
In the Liouville formulation the multicritical models correspond
the non-unitary conformal field theories coupled to 2d quantum gravity
and we will face the problem of correlation functions growing with distance.
This problem is currently under investigation.
In the case of unitary models coupled to gravity we expect
our philosophy to apply. These models usually have a representation
at the discretized level as short range interacting spin models
which at  specific temperatures become critical. For these models
there is a chance that estimates like \rf{6} might be valid.
The standard example is the Ising model on dynamical triangulations.
We hope to be able to solve this model by the technique outlined
above.

Finally it would be very interesting to generalize the above calculations
to higher dimensional simplicial quantum gravity. Work in this direction
is in progress \cite{aj1}.

\vspace{12pt}

\noindent
{\bf Acknowledgment} It is  a pleasure to thank Jerzy Jurkiewicz
for many interesting discussions.


\begin{thebibliography}{xx}

\bibitem{ddk} F. David, Mod.Phys.Lett. A3 (1988) 1651; J. Distler
              and H. Kawai, Nucl.Phys. B321 (1989) 509.

\bibitem{david} F. David, Nucl.Phys. B257 (1985) 45;
 Nucl.Phys. B257 (1985) 543.

\bibitem{adf} J. Ambj\o rn, B. Durhuus and J. Fr\"{o}hlich,
 Nucl.Phys. { B257} (1985) 433;  { B270} (1986) 457;
 { B275} (1986) 161-184.
\bibitem{adfo}
 J. Ambj\o rn, B. Durhuus  J. Fr\"{o}hlich and P. Orland,
  { B270} (1986) 457;  { B275} (1986) 161.

\bibitem{kkm}V.A. Kazakov, I. Kostov and A.A. Migdal, Phys.Lett. B157
(1985) 295; Nucl.Phys. B275 (1986) 641.

\bibitem{watabiki1}N. Kawamoto, Y. Saeki, and Y. Watabiki,
in preparation; see also
N. Kawamoto, INS-Rep.972;
Y. Watabiki, Prog.Theor.Phys.Suppl. No.114 (1993) 1.

\bibitem{transfer}H. Kawai, N. Kawamoto, T. Mogami and Y. Watabiki,
 Phys.Lett.B306 (1993) 19.

\bibitem{watabiki}Y. Watabiki, {\it Construction of Non-critical
String Field Theory by Transfer Matrix Formalism in Dynamical
Triangulation}, INS-Rep.1017, to be published in Nucl. Phys. B.

\bibitem{dfj}B. Durhuus, J. Fr\"{o}hlich  and T. Jonsson, Nucl.Phys.
B240 (1984) 453; B257 (1985) 779.

\bibitem{adj}J. Ambj\o rn, B. Durhuus and T. Jonsson,
Mod.Phys.Lett. { A6} (1991) 1133.

\bibitem{adj1}J. Ambj\o rn, B. Durhuus and T. Jonsson,
Phys.Lett. B244 (1990) 403.

\bibitem{aj} J. Ambj\o rn and J. Jurkiewicz,
Phys.Lett B278 (1992) 42.

\bibitem{am1} M. Agishtein amd A.A. Migdal, Mod.Phys.Lett A7 (1992) 1039.

\bibitem{david3}F. David, Mod.Phys.Lett. A5 (1990) 1019.

\bibitem{am}J. Ambj\o rn and Yu.M. Makeenko, Mod.Phys.Lett. { A5} (1990) 1753.


\bibitem{gk}S.S. Gubser and I.R. Klebanov, Nucl.Phys. B416 (1994) 827.

\bibitem{ackm}J. Ambj\o rn, L. Chekhov, C.F Kristjansen and Yu. Makeenko,
Nucl.Phys. B404 (1993) 127.

\bibitem{aj1} J. Ambj\o rn and J. Jurkiewicz, {\it Scaling in
four-dimensional simplicial quantum gravity}, to appear.

\end{thebibliography}
\end{document}